%% file: techRep.tex
\documentclass[12pt]{article}
\usepackage{epsfig,amssymb,amsmath}
\newcommand{\dt}{\partial_t}

\newcommand{\da}{\partial_a}

\newcommand{\Amax}{a_{\mbox{\tiny{max}}}}
\newcommand{\Amin}{a_{\mbox{\tiny{min}}}}

\newcommand{\Pmin}{p_{\mbox{\tiny{min}}}}

\newcommand{\ha}{\hat{a}}
\newcommand{\hth}{\hat{t}}
\newcommand{\hs}{\hat{\sigma}}
\newcommand{\hr}{\hat{r}}
\newcommand{\hxi}{\hat{\xi}}
\newcommand{\hmu}{\hat{\mu}}
\newcommand{\dht}{\partial_{\hth}}
\newcommand{\dha}{\partial_{\ha}}

\newcommand{\Phmin}{P_{\mbox{\tiny{min}}}}
\newcommand{\ahmin}{\hat{a}_{\mbox{\tiny{min}}}}
\newcommand{\ahmax}{\hat{a}_{\mbox{\tiny{max}}}}

\begin{document}
\def\trnum{2004-04}
\def\trauths{Bruce P. Ayati}
\def\trtitle{A Structured-Population Model of {\it Proteus mirabilis} Swarm-Colony Development}
\input{trcover}

\baselineskip=.33truein
\bigskip
\title{A Structured-Population Model of {\it Proteus mirabilis} Swarm-Colony Development}
\author{Bruce P. Ayati}

\maketitle

\medskip
\begin{abstract}
In this paper we present continuous age- and space-structured models
and numerical computations of {\em 
  Proteus mirabilis} swarm-colony development.  We base the
  mathematical representation of the cell-cycle
  dynamics of {\it Proteus mirabilis} on those developed by Esipov and
  Shapiro, which are the best understood aspects of the system, and we
  make minimum
  assumptions about less-understood mechanisms, such as precise
  forms of the spatial diffusion.  The models in this paper have
  explicit age-structure and, when solved numerically, display both
  the temporal and spatial regularity seen in experiments, whereas the
  Esipov and Shapiro model, when solved accurately, shows only the temporal
  regularity. 

The composite hyperbolic-parabolic partial differential equations used
  to model {\em Proteus mirabilis} swarm-colony development are
relevant to other biological 
  systems where the spatial dynamics depend on local
  physiological structure.  We use computational methods designed for such systems,
  with known convergence properties, to obtain the numerical results
  presented in this paper.  
\end{abstract}



\section{Introduction} 
\label{intro}

In this paper we present continuous age- and space-structured models
  and numerical computations of {\em 
  Proteus mirabilis} swarm-colony development. {\em
  Proteus mirabilis} has received treatment in the mathematical
  biology literature due to the strikingly regular
  spatial-temporal patterns swarm colonies make on an agar surface
  \cite{rauprich,WnSproteus}.   Not only do the swarm colonies form regularly
  spaced concentric terraces in a bulls-eye pattern, but the total
time taken to form a ring 
  is invariant under changes in the agar or glucose concentration in
  the substrate.

The invariance of the total cycle time rules out chemotaxis as an
explanation for the ring formation, as do biological observations of
{\em Proteus} responses to chemical stimuli \cite{WnSproteus}.  Esipov and Shapiro \cite{EnS} saw
that the regularity of the ring-formation cycle is a result of the
regularity of the differentiation-dedifferentiation cycle of
individual bacteria.  The resulting model was an equation that is
first-order hyperbolic in the age variable (also called a kinetic or
transport term) and nonlinear parabolic in space.

This mathematical explanation for swarm-colony behavior necessarily
differs from chemotaxis models, such as those for swarm-colony
formation in {\em E. coli}, due to the temporal invariance of the {\em
  Proteus} system \cite{Murray3-2}.

{\em Proteus} has been suggested as a model system for
problems in morphogenesis and other complex self-organizing systems
\cite{EnS,MnKnK,rauprich}, but it is also an example of a biological system where the
local kinetics -- the age structure -- determine the spatial dynamics
of the system.  {\em Dictostelium} has received much treatment as a model system for
other problems in morphogenesis \cite{MainiDicty}.  The study of {\em Proteus} can
have similar relevance to systems such as tumor growth \cite{cancerSurvey,DysonWebb,cancerModelling} and
natural forests
\cite{BolkerNPacalaNClaudia,Kohyama92FE,Kohyama93,Kalachev}, where
physiological structure, such as age, 
size or activation levels in tumor and immune cells, can play an important role in the spatial
behavior of the system.  Bertuzzi and Gandolfi \cite{BertuzziGandolfi}
presented a cancer tumor-cord model with both space and size
structure with some similarity to the kinetics in the {\em Proteus}
models, although they did not present results for their 
full model.

We base the models of {\it Proteus} swarm-colony development
in this paper on models presented in \cite{EnS}.   Due to the
phonetic and conceptual similarities of the terms ``swimmer'' and
``swarmer'', we depart from standard terminology and use the term
``dividing cell'' instead of ``swimmer cell''.  This change in
terminology emphasizes both the nonmotile character of the dividing cells
on an agar surface and the fact that the swarmer cells are
filaments\footnote{James Shapiro at the University of Chicago suggested
  this terminology when discussing the behavior of {\em Proteus} on an agar surface.}.

The models in \cite{EnS} indicated that experimental study of sharp
dedifferentiation in {\em Proteus} might yield insight into their periodic
swarm-colony development.  Dedifferentiation, also called septation,
is the process in which multinuclear swarmer cells subdivide into
mononuclear dividing cells.  This experimental study remains important
for the models in this paper, but perhaps more critical are
experiments that determine how
the lag phase in differentiation -- the process in which mitosis of dividing
cells results in a swarmer cell rather than two dividing cells --
affects swarm-colony development, and ideally give the functional forms of this
lag.  This lag in the onset of differentiation was identified in
\cite{WnSproteus}, and is an important part of the models in this
paper and of those in \cite{MnKnK}, but is not used in \cite{EnS}.  

We begin the paper with a description of {\em Proteus} swarm-colony
development and aspects of the relevant biology, and then discuss 
differences between the models in this paper and prior work. When
referring to the Esipov-Shapiro system, we mean a modified form of the
system formally presented in \cite{EnS}.  Without these changes, the
Esipov-Shapiro system has no swarming behavior at all and is thus not
really a model of the swarm-colony development.  In
Appendix \ref{appendixSergei}, we elaborate
upon the changes and why they might reflect the intended mechanisms of
Esipov and Shapiro.  Again, without these changes, we can have no useful
discussion of their model.

The most significant difference between the models in this paper and those in
\cite{EnS} is that the models in this paper reproduce the spatial
regularity seen in {\em Proteus} swarm
colony development, whereas, when solved accurately, the
Esipov-Shapiro system, even when modified with the maximum benefit of
doubt, does not.  The damping in terrace width is discussed in more detail in Appendix \ref{appendixSergei}.

In subsequent sections, we
present our mathematical models, the numerical methodology used to
solve the age-, space- and time-dependent systems, and the results from
the numerical computations.


\section{{\em Proteus mirabilis} Swarm-Colony Development}
\label{biology}

When inoculated onto an agar surface, {\em Proteus} begin formation of strikingly regular spatio-temporal
  patterns that begin with three initial phases: a lag phase, a first
  swarming phase, and a first consolidation phase; these form what
  appear to be the center circle and first ring of a bulls-eye.  This
  is followed by repeating cycles of ring, or terrace, formation that
  consist of a swarming phase followed by a consolidation
  phase.  These subsequent terraces are equal in width and time of
  formation.  Perhaps most interestingly, the time of terrace
  formation is invariant under changes in the glucose or agar
  concentration of the substrate, although changes in agar
  concentration do affect the ratio of time spent swarming versus
  consolidating, and changes in glucose increase terrace width.
  Terrace formation does vary by temperature \cite{rauprich}. 
 
Broth cultures of {\em Proteus} consist of mononuclear cells (often
called ``swimmer'' cells, but in this paper called ``dividing'' cells) about 0.6
$\mu$m wide and 1-2 $\mu$m long with short flagella.   On an agar
surface these cells lengthen somewhat to about 0.8
$\mu$m wide and 2-4 $\mu$m long.  After a period of time\footnote{This
  time-lag before swarmer cell production begins is incorporated 
  into the models in this paper and in \cite{MnKnK} as a lag in density, but is not used
  in \cite{EnS}.}, some cells
near the perimeter begin to lengthen into multinuclear filament cells (called
``swarmer cells'') that reach sizes of 0.7 $\mu$m wide and 20-80
$\mu$m long.  The swarmer cells have longer flagella than the
dividing cells.  Moreover, dividing cells normally have between 1 to
10 flagella, whereas swarmer cells have between 500 to 5000, a 10-fold
increase in the number of flagella per cubic micrometer of bacterial
volume \cite{WnSproteus}.  The process in which dividing cells become
swarmer cells is called ``differentiation.''  Differentiation only
occurs above one dividing-cells density and below another.

It is the swarmer cells that move across the agar during the
swarming phase.  The movement of swarmer cells on the agar surface is a complex
process. The swarmers use their elongated flagella to move in fluid
extracted by them from the agar gel.  They do so in groups, or
``rafts'' \cite{EnS}.  Consequently, the unit of random motion in any diffusion
approximation of {\em Proteus} movement is a large and
complicated entity. 

When the multinuclear swarmer cells approach a maximum size, they rapidly
break down into single-nucleus dividing cells.  This process is called
``dedifferentiation.''  Since mitosis results in exponential growth, the
size of a swarmer cell is an exponential weight times its age.
Thus, dedifferentiation occurs when swarmers reach a
maximum age.

Varying glucose concentration greatly alters biomass production
without greatly affecting the spatial or temporal aspects of the swarm
and the consolidation cycle of terrace formation.  Thus, the cycle is not
affected by biomass.  Changes in agar concentration do not change the
total time of terrace formation, but higher agar concentrations
shorten the time spent swarming and the width of the terraces, and
lengthen the time spent in consolidation \cite{EnS}.


\section{Differences and Commonalities with Previous Models}
\label{prior}

Models presented in this paper differ from models presented in \cite{EnS} in
several respects.  We use a radially symmetric geometry
with no-flux boundary conditions, as would be the case on a petri
dish.  Models used in \cite{EnS} use a linear geometry.  We use a diffusion approximation which
arises from isotropic random motion instead of Fickian diffusion
\cite{aronson85,IonidesStochastic,nagy89}.   An additional difference between the models in this paper
and in \cite{EnS} is that we present results for nondimensionalized models.

The most substantial differences between the
models presented here and in \cite{EnS} are in the different forms of the diffusivity,
$D$, and the differentiation rate to swarmer cells, $\xi$ (as will be discussed in
Section \ref{numerics}, our numerical treatments also differ, with
nontrivial consequences.)   The diffusivity in 
\cite{EnS} relied on a memory field: the diffusivity changes depending if the cells are in a collective state of
motion or at rest.  This is not clearly supported biologically and,
invoking Occam's Razor, is
not used in the models in this paper, since it is not needed.  Also,
the diffusivity in \cite{EnS} depends on the dividing-cell population.

The diffusivity in this paper depends only on swarmer biomass.
Since we use swarmer biomass rather than swarmer numerical density in
the diffusivity, the exponential weighting from mitosis that describes size means
that 
younger swarmer cells do not
contribute as much as older swarmer cells.  This weighting also
models the fact that each nucleus of a swarmer cell dedifferentiates into a dividing cell. Because of this weighting,
age-structured models of {\it Proteus} swarm-colony
development presented in this paper can also be thought
of as size-structured.  

The second and most important 
difference, the form of $\xi$, is based on incorporating the property
that dividing cells do not form swarmer cells until a minimum
population density is reached, a mechanism discussed in
\cite{WnSproteus} and
introduced in models in
\cite{MnKnK}, but not used in \cite{EnS}.  This threshold is represented 
mathematically by setting $\xi$ to zero for low dividing-cell
densities.

We focus on three features of {\em Proteus} swarm-colony
development: cycle regularity, ring-width uniformity, and control of
the ratio of swarm to consolidation time. A regular cycle does occur
in the Esipov-Shapiro system, and these authors are able to control the ratio
of swarm time to consolidation time.  However,
the terrace width decreases from cycle to cycle, so the front velocity
does not remain periodic, as shown in
Appendix \ref{appendixSergei}.

In \cite{EnS} it was concluded that models that used a sharp age of
septation from swarmer 
filament to dividing cells gave periodic front dynamics, whereas a 
constant dedifferentiation rate did not.

Medvedev, Kaper, and Kopell \cite{MnKnK} developed a reaction-diffusion model of
{\em Proteus} swarm-colony development based on averaging
over the age variable. The model relied on a specific nonlinear form
of the diffusivity to produce periodic front dynamics.   In \cite{MnKnK}, a constant
dedifferentiation rate was used rather than a sharp age of septation.
To obtain a periodic front velocity, the models in \cite{MnKnK} also introduced a
piecewise constant differentiation rate with a lag phase -- a biological mechanism
mentioned in \cite{WnSproteus} and used in the models in this paper --
and a diffusivity dependent on  
the ratio of swarmer-cell to dividing-cell biomass.  The use of a
constant dedifferentiation rate allowed the age-structured
diffusion system to be reduced to a 
reaction--diffusion system.  This reduction has advantages both
numerically and analytically.  Moreover, a reaction-diffusion equation
with periodic front dynamics is an interesting mathematical object in
its own right.  However, removing explicit age eliminated a
mechanism for controlling the ratio of time spent swarming to time
spent in consolidation without changing the total cycle time.  No
mechanism within the reaction-diffusion framework was suggested to
replace this.

The biological mechanisms
that are best understood are emphasized in the models in this paper.  In
particular, we maintain the explicit age structure. Due to
experimental results\footnote{Preliminary experiments conducted by
  James Shapiro's laboratory at the University of Chicago suggest that {\em Proteus 
    mirabilis} that have been genetically modified to no longer have a
  sharp septation age lose periodicity in their swarm-colony
  development.} which postdate previous work \cite{EnS,MnKnK}, we are
more confident in assuming a sharp age of septation.  New
numerical methods for handling age-structured population models with
nonlinear diffusion have
been developed so that computations that incorporate the
known properties of {\it Proteus} swarm-colony
development can be run efficiently
\cite{age-pwconst-paper,age-general-paper}.  

Because less is known
about the specific forms of the diffusion and diffusivity, we attempt
to use the simplest form of diffusivity suitable for this system.
Rather than using a relatively complicated fractional form involving
both swarmer and dividing cells, we use a diffusivity that is
proportional to swarmer biomass above a threshold, reflecting the need
for swarmers to group together in order to move.  This is the simplest
form that is biologically justified and that results in both swarming and
consolidation.  This density dependence on
swarmers, due to ``raft'' building, is not unique to the models in
this paper and is also used in
\cite{EnS} and \cite{MnKnK}.  The diffusivity we use, since $\xi$ is
zero above some threshold, also results in an absence of swarmers in
the interior of the colony -- the primary justification for the
dependence of the diffusivity on dividing cells in the other models.

The different model mechanisms and results for those in this paper,
Esipov-Shapiro, and Medvedev et al. are summarized in Table
\ref{summary}.  The models in this paper are unique in reproducing
three major aspects of {\em Proteus} swarm-colony development:
temporal regularity of overall cycle time,
  spatial regularity, and control of the ratio of swarm time to
  consolidation time.  Moreover, by retaining the explicit
  age structure in the mathematics, the models reproduce the colony behavior by
  relying on only two main mechanisms.

\begin{table}
\begin{center}
\begin{tabular}{|l||c|c|c|}
\hline
Mechanism $\backslash$ Model & Ayati & Esipov-Shapiro & Medvedev et al. \\ \hline\hline
Dependence of diffusivity & & & \\
on swarmer density & $\surd$ & $\surd$ & $\surd$ \\ \hline
Dependence of diffusivity & & & \\
on dividing-cell density & & $\surd$ & $\surd$ \\ \hline
Memory field & & $\surd$ &  \\ \hline
Lag phase in $\xi$ & $\surd$ &  & $\surd$ \\ \hline
Explicit age structure & $\surd$ & $\surd$ &\\ \hline\hline
Behavior $\backslash$ Model & Ayati & Esipov-Shapiro & Medvedev et al. \\ \hline\hline
Regularity of swarm and &&&\\ 
consolidation cycle & $\surd$ & $\surd$ & $\surd$ \\ \hline
Regularly spaced &&&\\ 
concentric terraces & $\surd$ &  & $\surd$ \\ \hline
Control of ratio of swarm  &&&\\ 
to consolidation time & $\surd$ & $\surd$ & \\ \hline\hline

\end{tabular}
\caption{Mechanisms and behaviors for different models of {\em Proteus
    } swarm-colony development.  The models in this paper are
    unique in reproducing three major aspects of the colony
    development, and, by retaining the explicit
  age structure, reproduce the colony behavior with only two main mechanisms.}
\label{summary}
\end{center}
\end{table}


\section{The Model} 
\label{model}

The models in this paper and in \cite{EnS} are based on a mathematical
framework with some history behind it.  Skellam \cite{skellam}
considered the effects of diffusion on populations in his classic work
of 1951.  Sharpe and Lotka in 1911 and McKendrick in 1926 considered population models with linear age
structure \cite{Webb}. More recently, Gurtin and MacCamy \cite{GnM1}
considered models with nonlinear age structure.  Rotenberg
\cite{roten} and  Gurtin \cite{gurtin} posed models dependent on both
age and space.  Gurtin and MacCamy \cite{GnM2} differentiated between
two kinds of diffusion in these models: diffusion due to random
dispersal, and diffusion toward an area of less crowding.  Existence
and uniqueness results can be found for various forms of these models
in  Busenberg and Iannelli \cite{BnI}, di Blasio \cite{diblasio}, di
Blasio and Lamberti \cite{DnL}, Langlais \cite{langlais1}, and MacCamy
\cite{maccamy}. Further analysis has been done by several authors
\cite{huang,KnL,langlais2,marcati}. 

For our model, we assume the colony is radially symmetric.  The
variables $r$, $a$, and $t$ represent radius in two-dimensional space,
age, and time, respectively.
The function $u(r,a,t)$ represents the swarmer-cell population density at
radius $r$, age $a$, and time $t$.  The functions $p(r,t)$ and $v(r,t)$
represent the biomass density of swarmer cells and the dividing-cell population
density, respectively, at radius $r$ and time $t$.  The dividing-cell
population density, $v$, is 
measured in population-density units (pdu) with $\mbox{pdu} =
\mbox{Gc}/\mbox{cm}^2$, where Gc denotes one billion cells. 
The swarmer-cell population density, $u$, is measured in pdu/hr.  The
swarmer-cell biomass density, $p$, is measured in $\mbox{mg}/\mbox{cm}^2$.  

The dimensional model consists of the age-structured nonlinear diffusion system 
\begin{subequations}
\begin{eqnarray}
\dt u + \da u &=& \frac{1}{r}  \partial_r \big( r \partial_r ( D(p) u)
\big) - \mu(a)u, \  0 \leq r < r_m, \, a>0, \, t > 0,
\label{swarm} \\
\dt v &=&  \frac{1 - \xi(v)}{\tau}v + \int_0^\infty \mu(a) \, u \, e^{a/\tau} \ da, 0 \leq r \leq r_m,  t > 0, \label{swim} \\
u(r,0,t) &=& \frac{\xi(v)}{\tau}v(r,t),  0 \leq r \leq r_m, \ t > 0
\label{birth}
\end{eqnarray}
with boundary and initial conditions
\begin{eqnarray}
\partial_r \big( D(p(r_m,t)) u(r_m,a,t) \big) = 0,&&  a > 0 \ ,  t > 0, \\
u(r,a,0) = 0, &&  0 \leq r \leq r_m, a \geq 0,
\label{initial-u}\\
v(r,0) = v_0(r),&&  0 \leq r \leq r_m. \label{initial-v}
\end{eqnarray}
The parameter $\tau$ is the time it takes a cell to subdivide. A typical value of $\tau$ is 1.5 hours.  The
total motile swarmer cell biomass is given by 
\begin{equation}
p(r,t) =  m_0 \int_{\Amin}^{\infty} u(r,a,t) e^{a/ \tau} \ da,  \qquad  0 \leq r \leq r_m, \ t\geq 0, \label{Ps}
\end{equation}
where $m_0$ is measured in mg/Gc.  A
dividing cell has a volume of approximately $2\times
10^{-18}\mbox{m}^3$ so that $m_0\approx 2$ for {\it Proteus mirabilis}.

{\em Proteus} move through a process of raft building that requires
two things: sufficient maturity in swarmer cells to contribute
to raft building, and a sufficient biomass of mature cells to form the
rafts. The lower limit of integration, $\Amin$, is the minimum age when a
swarmer cell is sufficiently large, with sufficiently long flagella,
to contribute to motion on the agar surface.  The parameter $\Amin$ is
comparable to the parameter $\theta_{\mbox{\tiny{min}}}$, the lower limit of
integration in the total swarmer numerical density, in \cite{EnS}.  Each
parameter controls the ratio of swarm time to consolidation
time without changing the total time of ring formation.  The parameter
$\Amin$ is related to agar concentration.  The higher the
concentration, the drier the surface, raising the value of $\Amin$.
Higher agar concentration, and thus higher $\Amin$, shortens
the swarming phase and lengthens the consolidation phase in terrace formation.

The diffusivity has the form 
\begin{equation}
  D(r,t) = D_0 \max\big\{ (p(r,t)-\Pmin), 0\big\}. \label{diffusivity}
\end{equation}
The parameter $\Pmin$ is the minimum biomass needed for swarmers to
build rafts capable of moving on the agar surface, and is roughly
equivalent to $P_{s,\mbox{\small min}}$ in \cite{EnS}.

We use a differentiation function with a lag phase that is a $C^1$ piecewise cubic with
support of length $2v_w$: 
\begin{equation}
 \xi(v) =  \left\{ \begin{array}{ll} \xi_0 \left( 2\left(
                      \frac{|v-v_c|}{v_w} \right)^3 - 3\left(
                      \frac{v-v_c}{v_w} \right)^2 + 1 \right), &
                      |v-v_c| \leq v_w, \\
                      0, & \mbox{otherwise,}
                      \end{array} \right.  \\ \label{diff}
\end{equation} 
The interval $[v_c-v_w,v_c+v_w]$ is the
swarmer-cell production window.  The parameter $v_c$ represents the
lag phase in swarmer-cell production \cite{MnKnK}.  We use a compactly
supported cubic function because it is smooth.  The important aspects
of the functional form of $\xi$ are that it is zero above and below
certain thresholds.  Appendix \ref{appendixXi} shows
that the nature of the results do not depend strongly on the shape of
the curve.  The shape of the curve does help with numerical issues
concerning the degenerate diffusion.  The parameter $v_c$ has no
analog in \cite{EnS}.

The dedifferentiation modulus, with a spread parameter $\sigma$, is
\begin{equation}
 \mu(a) = \frac{\mu_1}{\sigma} \exp\big( (a-\Amax)^2/(2 \sigma^2)
 \big) + \frac{\mu_2}{\sigma} H(a), \label{mu}
\end{equation}  
where 
$$ H(a) = \left\{ \begin{array}{ll}
 1, & \ a \geq 0, \\
 0, & \ a < 0 \end{array} \right. $$
is the Heaviside function.  This dedifferentiation modulus represents
 a situation where the probability of dedifferentiation is low for
 young swarmers, increases as they approach $\Amax$ in age, and
 remains high afterwords.  Experiments indicate that this type of
 dedifferentiation is dominant for {\em Proteus mirabilis}, and that
 the transition, from virtually no probability of dedifferentiation to
 an extremely high probability of dedifferentiation as a function of age, occurs very
 rapidly near the critical age, $\Amax$.  Preliminary results from
 James Shapiro's laboratory at the University of Chicago show that
 different strains of {\em Proteus} with less sharp dedifferentiation
 have oscillatory, but not strictly periodic, front velocities.

We take the initial condition to be 
\begin{equation}
 v_0(r) = \left\{ 
                \begin{array}{ll}
                      v_h \left( 2\left( \frac{r}{r_0} \right)^3
                            - 3\left( \frac{r}{r_0} \right)^2 + 1 \right), & 0 \leq r \leq r_0, \\
                      0, & r > r_0.
                \end{array} 
          \right.
\label{initial-form}
\end{equation} 
\end{subequations}
Again, it is mass, not shape, that matters qualitatively. Shape effects numerical efficiency.

\noindent In the limit $\sigma \rightarrow 0$, which represents a
sharp age at which all dedifferentiation occurs, we have a system on a
bounded age domain  
\begin{subequations}    
\begin{eqnarray}
\dt u + \da u &=& \frac{1}{r}  \partial_r \big( r \partial_r ( D(p) u)
\big), \ 0\leq r < r_m,  0< a < \Amax,  t > 0, \\
\dt v &=&  \frac{1 - \xi(v)}{\tau}v + u(r,\Amax,t)e^{\Amax/\tau}, \   0\leq r \leq r_m, \ t > 0, \\
p(r,t) &=&  m_0 \int_{\Amin}^{\Amax} u \, e^{a/ \tau} \ da, \ 0\leq r \leq r_m, \ t\geq 0
\end{eqnarray}
with conditions
\begin{eqnarray}
u(r,0,t) &=& \frac{\xi(v)}{\tau}v(r,t), \ 0\leq r \leq r_m, \ t > 0, \\
\partial_r \big( D(p(r_m,t)) u(r_m,a,t) \big) &=& 0,  \ 0 < a < \Amax, \ t > 0, \\
u(r,a,0) &=& 0,   \ 0\leq r \leq r_m, 0 \leq a < \Amax, \\
v(r,0) &=& v_0(r),   \ 0\leq r \leq r_m,
\end{eqnarray} 
\end{subequations}

\subsection{A Nondimensional Model} 
\label{nondim}

We nondimensionalize Equations (\ref{swarm})-(\ref{initial-form}).  We scale age and time by $\tau$ and
space by $r_m$ and introduce the independent variables  

$$\hr = r/r_m,\; \ha = a/\tau,\; \hth = t/\tau.$$  

\noindent We scale the dependent variables,
 
$$U(\hr,\ha,\hth) = u(r,a,t)/(v_w/\tau),\; V(\hr,\hth) = v(r,t)/v_w,\; P = p/(m_0 v_w).$$

\noindent We use the dimensionless parameters
$$\hat{D}_0 = D_0 m_0 v_w \tau/r_m^2,\; \Phmin = \Pmin/(m_0 v_w),\; V_c = v_c/v_w,$$ 
$$\ahmax = \Amax/\tau,\; \ahmin = \Amin/\tau,\; \hs = \sigma/\tau.$$

We note that in this nondimensionalization, the length of the swarmer-cell production
window (2$v_w$) and swarmer-cell swarm threshold ($\Pmin$) are
related.  It seems intuitive that halving one is equivalent to
doubling the other.    The effect of a lag in swarmer cell production
($v_c$) is isolated from the effects of $v_w$ and 
$\Pmin$.  \\

\noindent For $\hs \neq 0$, we obtain the system  
\begin{subequations}   
\begin{eqnarray}
\dht U + \dha U &=& \frac{1}{\hr}  \partial_{\hr} \big( \hr \partial_{\hr} (\hat{D}(P) U)
\big) - \hmu(\ha)U, \  0\leq r < 1, \ \ha > 0, \ \hth > 0, \\
\dht V &=&  \big(1 - \hxi(V)\big) V + \int_0^\infty \hmu(\ha) \, U \, e^{\ha} \ d\ha,\ 0\leq \hr \leq 1, \ \hth > 0,\\
P(\hr,\hth) &=& \int_{\ahmin}^{\infty} U(\hr,\ha,\hth) e^{\ha} \ d\ha,\  0\leq \hr \leq 1, \ \hth\geq 0
\end{eqnarray}
with conditions
\begin{eqnarray}
U(\hr,0,\hth) = \hxi(V) V(\hr,\hth), &&  0\leq \hr \leq 1, \ \hth > 0, \\
\partial_{\hr} \big( \hat{D}(P(1,t)) U(1,\ha,\hth) \big) = 0,&&  \ha > 0,  \hth > 0, \\
U(\hr,\ha,0) = 0, &&  0\leq \hr \leq 1, \ \ha\geq 0\\
V(\hr,0) = v_0(r_m \hr)/v_w,&&  0\leq r \leq r_m,
\end{eqnarray}

\noindent where
\begin{eqnarray}
  \hat{D}(\hr,\hth) &=& \hat{D}_0 \max\big\{ (P(\hr,\hth)-\Phmin), 0\big\}, \\
 \hxi(V) &=&  \left\{ \begin{array}{ll} \xi_0 \left( 2|V-V_c|^3 - 3(V-V_c)^2 +
                      1 \right), &  |V-V_c| \leq 1, \\ 
                      0, & \mbox{otherwise,}
                      \end{array} \right.  \\ \label{diffU}
 \hmu(\ha) &=& \frac{\hmu_1}{\hs} \  \exp\big( (\ha-\ahmax)^2/(2 \hs^2)
 \big) +  \frac{\hmu_2}{\hs}H(\ha).
\end{eqnarray}
\end{subequations} 
 
\noindent In the limit $\hs \rightarrow 0$ we have the system   
\begin{subequations} 
\begin{eqnarray}
\dht U + \dha U &=& \frac{1}{\hr}  \partial_{\hr} \big( \hr \partial_{\hr} ( \hat{D}(P) U)
\big),\  0\leq \hr <1, \, 0< \ha < \ahmax, \, \hth > 0, \label{U} \\
\dht V &=&  (1 - \hxi(V))V + U(\hr,\ahmax,\hth)e^{\ahmax}, \ 0 \leq \hr \leq 1, \ \hth > 0, \label{V} \\
P(\hr,\hth) &=& \int_{\ahmin}^{\ahmax} U \, e^{\ha} \ d\ha, \ 0 \leq \hr \leq 1, \ \hth\geq 0, \label{P}
\end{eqnarray}
with conditions
\begin{eqnarray}
U(\hr,0,\hth) = \hxi(V)V(\hr,\hth), && 0 \leq \hr \leq 1, \ \hth > 0,
\label{Ubirth} \\ 
\partial_{\hr} \big( \hat{D}(P(1,\hth)) U(1,\ha,\hth) \big) = 0, && 0< \ha < \ahmax, \ \hth > 0, \\
U(\hr,\ha,0) = 0, && 0 \leq \hr \leq 1, \ 0< \ha < \ahmax, \label{initial-U} \\
V(\hr,0) = v_0(r_m \hr)/v_w, && 0 \leq \hr \leq 1. \label{initial-V}
\end{eqnarray}
\end{subequations} 

In this paper we will examine only the limiting case of $\hs \rightarrow 0$, which
corresponds to the case of a sharp age of septation, ``Model A'',
in \cite{EnS}.  The models used in the computations in this paper are
nondimensional, whereas dimensional models were used in \cite{EnS}.

As mentioned earlier, there are two major differences between the models in
this paper and those in \cite{EnS}.  First, the diffusivity is of much simpler
form, with no memory field, and hence no parameter for the upper
threshold for diffusion in a memory field ($P_{\mbox{\tiny{max}}}$).
Second, there is a lag 
phase in the dedifferentiation function, $\xi$.  Also, we use
biomass, not numerical density (which 
was how the model was stated in \cite{EnS}) as the relevant measure of
the totality of swarmers.  Other differences
include radial symmetry, and diffusion based on
isotropic random motion rather than Fickian diffusion. 

When solved accurately, these differences yield the temporal regularity seen in
{\em Proteus} experiments, as well as the regularly spaced
concentric terraces, something not seen in accurate computations of the Esipov-Shapiro system.


\section{Numerical Methodology}
\label{numerics}

There has been much investigation into numerical methods for solving
models with just age structure \cite{chiu,FnL-M,KnC,L-M,sulsky}. 
Kim \cite{kim}, Kim and Park \cite{KnP}, and Lopez and Trigiante \cite{LnT} developed
methods for age-structured populations that undergo random dispersal
in space.  All these methods involve uniform time and age
discretizations, with the age step chosen to equal the time step.
These methods discretized along
characteristics, but they did so simultaneously in age and time and
thus imposed the often 
crippling constraint that the time and age steps must be both constant and
equal.   The difficulty with this approach is twofold.  First, the
use of constant age and time
steps prevents adaptivity of the discretization in age or, especially,
time.   Second, and more 
important, the coupling of the age and time meshes can cause great
losses of efficiency, since only rarely will the dynamics in time be on
the same scale as the dynamics in age.  This is particularly the case
when space is involved, since sharp moving fronts can require small
time steps, 
whereas the behavior in the age variable can remain relatively smooth.

The
methods used in this paper \cite{age-pwconst-paper,age-general-paper}
use a moving age discretization that allows for nonuniform 
age and time discretizations. The age step need not equal the time
step. Instead, the positions of the age nodes are adjusted by the time
step. The methods preserve the important fact that age and time
advance together.  The age discretizations presented in \cite{kim,KnP,LnT} can be viewed
as special cases of the methods 
presented in \cite{age-pwconst-paper,age-general-paper} by setting the
time and age 
meshes to be constant and equal and using a backward Euler
discretization in time and a piecewise constant finite element space
in age.

The importance of allowing different age and time discretizations is
somewhat illustrated by the application of the methods of de Roos
\cite{deRoos89}.   These methods have found use in the study of
ecological systems such as {\it Daphnia} (see \cite{deRoos97} and the
references therein) as well as in theoretical population biology
\cite{KooiNKooijman,PascualNLevin}.  The methods of de Roos are
formulated for the 
case of time and a variable representing some sort of physiological
structure, most simply age, and involve
moving the age nodes along characteristics.  They have not been
formulated for explicit space.
The representation of the approximate solution is probabilistic
and not functional, and birth and death are handled differently than in
the methods in \cite{age-pwconst-paper,age-general-paper}.  Even so,
it would be interesting to know 
if an energy analysis 
could provide a framework for the convergence analysis sought in
\cite{deRoos91}.    The main effect of de Roos's
methods is to separate the age and time discretizations, thereby
yielding an approximation that is dispersion free in age, in order to
provide a method that works in practice.

We emphasize the importance of reliable numerics.   In the systems
presented in this paper and in \cite{EnS}, the regularity in terrace
formation is due to the regularity in the aging of the swarmer cells.
A coarse, stage-structured numerical approximation to the continuous
aging term, such as that used in \cite{EnS}, can constitute a
qualitatively different swarmer-cell life cycle than that of the
original continuous model. Also, the degenerate diffusion can be
altered by the choice of the time step.  The numerical computation may then show
periodic front dynamics that are not obtained by an accurate solution
of the original continuous equations, but are induced by a regularity
in the numerics.  Appendix \ref{appendixSergei} details
how accurate computations change the behavior of the
Esipov-Shapiro system from what is presented in \cite{EnS}.

{\em Proteus mirabilis} swarm-colony development is important as a
model system for other problems in morphology, complex
self-organizing systems, and where spatial behavior is a manifestation
of local kinetics.  Not only are the choices made in the mathematical
models important in this respect, but so is the computational framework
used to solve them.  One goal of this paper is to illustrate the
necessity of robust and accurate new numerical methods in the modeling
and solution of such systems.

The results of the numerical computations presented in Section
\ref{results} were obtained for the various parameter sets by using
the age and space discretizations described in
\cite{age-pwconst-paper,age-general-paper} with time integration done
using step-doubling extrapolation.
Step-doubling is a time-stepping algorithm, going back to at least
Gear \cite{gear} for ordinary differential equations, based on taking one step
over a time interval to obtain one approximate solution for that
interval, and then taking two half-steps over the same interval to get
a second approximate solution.  The two solutions can then be compared
for adaptive step-size control.  Moreover, the two solutions can be combined to
obtain a likely second-order approximation in time
\cite{step-doubling-paper}.  The use of a second-order adaptive method
with error checking for these problems is important, since the front
dynamics of a degenerate system can depend on the time step.

A uniform spatial discretization of size $\Delta x = 1/300$, and a
uniform age discretization of $\Delta a = 1/40$, using piecewise
constant basis functions, was sufficient to
solve the system within a relative error in the $L^2$-norm of less than
1\%.  A uniform age discretization in the context of a moving grid
method means that all but the first and last age intervals are
constant in length, and that a new age interval is introduced at the
birth boundary when the old birth interval reaches $\Delta a$ in length.
Convergence in time was obtained by adjusting a tolerance parameter
for the adaptive time-stepping in the step-doubling algorithm so that
the relative error in the $L^2$-norm was also less than 1\%. 


\section{Results}
\label{results}

\begin{figure}
\centering
\epsfig{file=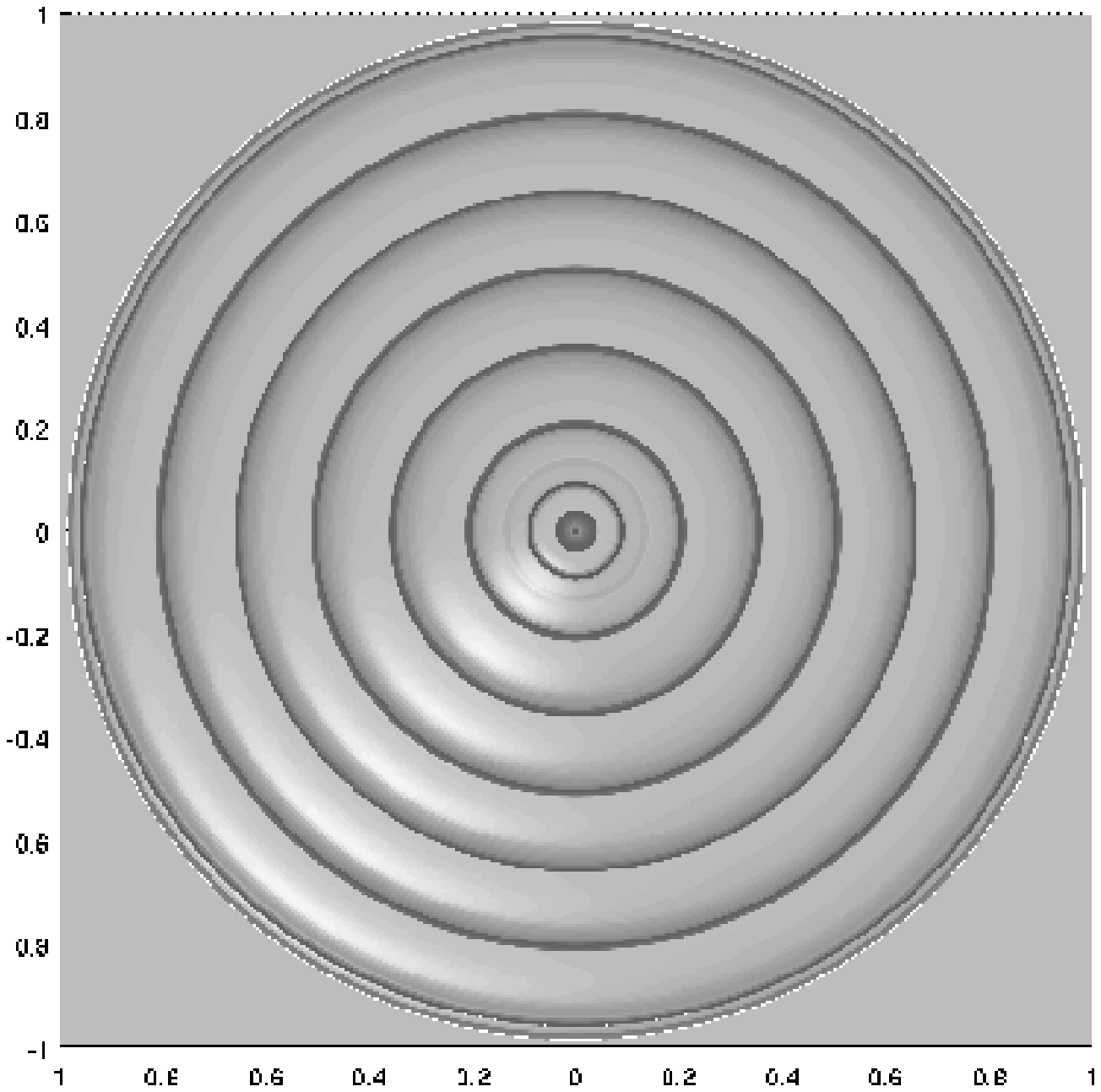,height=3in} \\ \vspace{0.1in}
\epsfig{file=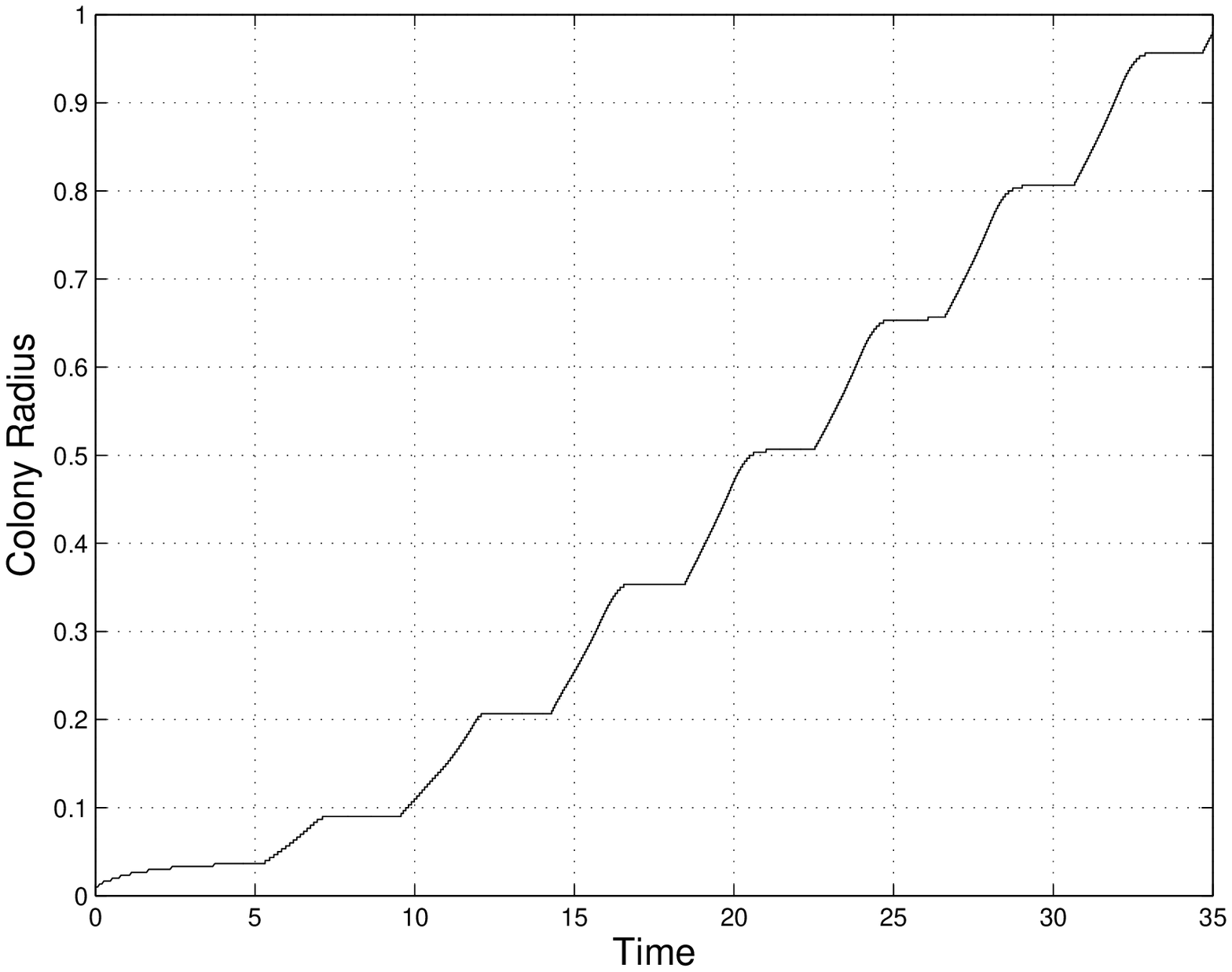,height=2.7in}
\caption{Computed {\em Proteus mirabilis} swarm colony with 
  regularly spaced concentric terraces and periodic front velocity.
  The parameters for this swarm colony are $\ahmax = 2.67$, $V_c = 8$,
  $D_0 = 2\times 10^{-3}$, $\Phmin = 0.5$, $\xi_0 = 0.5$, and $\ahmin=0$.
  The top figure is a 3D plot of
  the swarmer- and dividing-cell biomass viewed from directly above
  with a combination of ambient light, diffuse reflection, and specular
  reflection. The bottom figure shows the radius of the swarm
  colony as a function of time.  Areas where the function has zero
  slope correspond to consolidation periods in the swarm-colony
  development.  In dimensionless units, total time for the formation
  of one ring is $T = 4.1$, the time spent swarming is $S =2.2$, the
  time spent in consolidation is $C = 1.9$, and the width of one ring
  is $R = 0.153$.}
\label{baseColony} 
\end{figure}

We examine the response of the model to changes in the various
 parameters in the following tables.  The spatial domain of the problems is $\hr \in [0,1]$, the temporal
domain is truncated to the finite domain $\hth \in [0,35]$, and the age
domain is $\ha \in [0,\ahmax]$, which varies depending on the age of
sharp septation in swarmers.

 The quantity $T$ is the total time it takes for a swarm ring to form,
  $S$ is the time spent in a swarm phase, $C$ is the time spent in a
  consolidation phase, and $R$ is the width of a swarm ring.  Except
 for the parameter under study, the other parameters are set to
 $\ahmax = 2.67$, $V_c = 8$, $D_0 = 2\times 10^{-3}$, $\Phmin = 0.5$,
 $\xi_0 = 0.5$, and $\ahmin = 0$.  In each of the tables, the default value of the
 parameter will be shown in bold type.  Figure \ref{baseColony}
 shows the results of the model for the base parameters.  The top
 graph is a three dimensional plot of the physical colony viewed from
  above and the bottom graph is
 a plot of colony radius versus time that illustrates the periodic
 front dynamics of {\em Proteus} swarm-colony development.
 
\begin{table}
\begin{center} 
\begin{tabular}{|r||r|r|r|r|r|r|}
\hline
$\ahmin$ & $T$ & $S$ & $C$ & $S/C$ & $R$ & $R/S$ \\ \hline
{\bf 0.0}&4.1&2.2&1.9&1.16&0.153&0.0695 \\ \hline
0.5&4.1&2.0&2.1&0.95&0.150&0.0750 \\ \hline
1.0&4.1&1.7&2.4&0.71&0.140&0.0824 \\ \hline
2.0&4.1&1.3&2.8&0.46&0.090&0.0692 \\ \hline
\end{tabular}
\caption{Ring formation as a function of $\ahmin$.}
\label{ahMin}
\end{center}
\end{table}

Table \ref{ahMin} exhibits ring formation as a function of $\ahmin$, the
minimum nondimensional age for a swarmer cell to contribute to
swarming.  The parameter $\ahmin$ depends on agar
concentration.  As agar concentration increases, so does the
difficulty of moving on the substrate, and hence $\ahmin$ increases as well.  As $\ahmin$
increases, $S$ and $R$ decrease, $C$ increases, and $T$ stays the same.
This is the response to increased agar concentrations seen in experiments;
as agar concentration increases, swarm time and terrace width decrease, consolidation
time increases, whereas total cycle time stays the same.  

\begin{figure}[t]
\centering
\epsfig{file=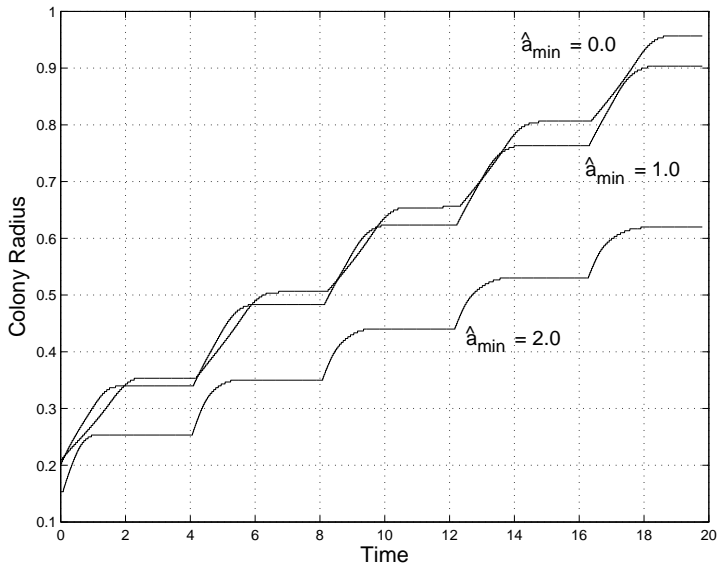,height=3in} 
\caption{Swarm-colony front dynamics as a function of $\ahmin$.  Shown
  are curves for the cases of $\ahmin = 0.0$, 
  $\ahmin = 1.0$,  and $\ahmin = 2.0$.
  The other parameters for these swarm colonies are $\ahmax = 2.67$, $V_c = 8$,
  $D_0 = 2\times 10^{-3}$, $\Phmin = 0.5$, and $\xi_0 = 0.5$.  The
  curves have been realigned in time for better
  comparison of individual ring formation.  Although total cycle time
  does not change, the ratio of time spent swarming to time spent in
  consolidation decreases as the minimum nondimensionalized age to
  contribute to swarming, $\ahmin$, increases.} 
\label{aMin}
\end{figure}

This invariance of the total time for ring formation, $T$, is further
illustrated in Figure \ref{aMin}.  A parameter similar to $\ahmin$
controls the ratio of swarm time to consolidation time in \cite{EnS}.
Having averaged out the age variable, no such control is shown in
\cite{MnKnK}. 

\begin{table}
\begin{center}
\begin{tabular}{|r||r|r|r|r|r|r|}
\hline
$D_0$ & $T$ & $S$ & $C$ & $S/C$ & $R$ & $R/S$ \\ \hline
0.001&4.05&2.15&1.9&1.13&0.107&0.0498\\ \hline
{\bf 0.002}&4.1&2.2&1.9&1.16&0.153&0.0695\\ \hline
0.003&4.1&2.3&1.8&1.28&0.183&0.0796\\ \hline
0.004&4.1&2.3&1.8&1.28&0.214&0.0930\\ \hline
\end{tabular}
\caption{Ring formation as a function of $D_0$.}
\label{D0}
\end{center}
\end{table}

Table \ref{D0} exhibits ring formation as a function of $D_0$, the
constant of diffusion.  The parameter $D_0$ reflects changes in
glucose concentration.  The 
temporal metrics, $T$, $S$, and $C$, change little, whereas there is
significant change in the width of the terraces, $R$.  This
corresponds to the experimental results illustrated in Figure 14 in
\cite{rauprich}, which shows minor variation in the total cycle
time and the ratio of swarm to consolidation time, while showing a
significant increase in terrace width.

\begin{table}
\begin{center}
\begin{tabular}{|r||r|r|r|r|r|r|}
\hline
$V_c$ & $T$ & $S$ & $C$ & $S/C$ & $R$ & $R/S$ \\ \hline
2   &3.3&2.1&1.2&1.75 &0.107&0.0510  \\ \hline
4&3.6&2.2&1.4&1.57&0.133&0.0605 \\ \hline
{\bf 8}&4.1&2.2&1.9&1.16&0.153&0.0695 \\ \hline
16&4.5&2.2&2.3&0.96&0.157&0.0714 \\ \hline
\end{tabular}
\caption{Ring formation as a function of $V_c$.}
\label{Vc}
\end{center}
\end{table}

Table \ref{Vc} exhibits ring formation as a function of $V_c$, the
nondimensionalized lag in the swarmer cell production, $\xi$.
As $V_c$ increases, we see an expected increase in the consolidation
time, $C$.  The time spent swarming, $S$, is nearly invariant.  This leads to an increase in the overall cycle time,
$T$, as $V_c$ increases.  The increased time building up swarmer rafts
results in an increase in the front velocity, $R/S$, and thus in
terrace width, $R$, as well. 

The presence of a lag in swarmer-cell production was observed in
\cite{WnSproteus}. 
However, the length of the lag, much less its explicit density dependence,
is an open area of experimentation.  The biological and chemical
factors that underlie $V_c$ are also unknown. 

\begin{table}
\begin{center} 
\begin{tabular}{|r||r|r|r|r|r|r|}
\hline
$\xi_0$ & $T$ & $S$ & $C$ & $S/C$ & $R$ & $R/S$ \\ \hline
0.25&4.7&1.65&3.05&0.54&0.067&0.0406\\ \hline
{\bf0.50}&4.1&2.2&1.9&1.16&0.153&0.0695\\ \hline
0.75&3.7&2.6&1.1&2.40&0.230&0.0885\\ \hline
1.00& N/A & N/A & N/A & N/A & N/A & N/A \\ \hline
\end{tabular}
\caption{Ring formation as a function of $\xi_0$.}
\label{xi0}
\end{center}
\end{table}

Table \ref{xi0} exhibits ring formation as a function of $\xi_0$, the
constant of the differentiation ratio, $\xi$.  The parameter $\xi_0$
is clearly greater than zero, otherwise dividing 
cells would never beget swarmers, and is thought to be less than one,
as suggested in \cite{WnSproteus}, ``some of the cells near the
perimeter of the colony begin to undergo a dramatic morphological
change to highly elongated forms$\dots$''  Results for $\xi_0 = 1$
show the loss of a swarming consolidation cycle and a nearly constant
rate of colony expansion.

As $\xi_0$ increases, the consequent increase in swarmer-cell
production results in longer time spent swarming, $S$, and terrace
width, $R$, but a decrease in total cycle time, $T$, and consolidation
time, $C$.  

\begin{table}
\begin{center}
\begin{tabular}{|r||r|r|r|r|r|r|}
\hline
$\Phmin$ & $T$ & $S$ & $C$ & $S/C$ & $R$ & $R/S$ \\ \hline
0.0&4.3& N/A & N/A & N/A &0.186& N/A \\ \hline
0.2&4.2&2.8&1.4&2.00&0.170&0.0607\\ \hline
0.3&4.15&2.6&1.55&1.68&0.163&0.0627\\ \hline
{\bf 0.5}&4.1&2.2&1.9&1.16&0.153&0.0695\\ \hline
0.7&3.95&2.05&1.9&1.08&0.140&0.0683\\ \hline
1.0&3.85&1.9&1.95&0.97&0.127&0.0668\\ \hline
\end{tabular}
\caption{Ring formation as a function of $\Phmin$.}
\label{Phmin}
\end{center}
\end{table}

Table \ref{Phmin} exhibits ring formation as a function of $\Phmin$, the
minimum swarmer biomass for swarming to begin.  The parameter $\Phmin$
reflects the need for swarmers to build 
``rafts'' with other swarmers in order 
to move. ``Movement of individual cells seems to be retarded whereas
cells moving in larger groups do so much more effectively''
\cite{WnSproteus}.  Rafts are able to form in sufficient numbers only when there
is a minimum swarmer biomass present.  The parameter $\Phmin$ differs
from $\ahmin$ in that it is a measure of weighted population density
rather than of the size of an individual.  It is unclear what biological
or chemical factors set the minimum swarmer biomass, or even if this
threshold differs substantially as experimental parameters vary.  

As $\Phmin$ increases, the total time spent swarming, $S$, decreases,
and the time spent in consolidation, $C$ increases.  This is the
same as the response of the system to changes in $\ahmin$, except that the
total cycle time, $T$, remains strongly invariant under changes in $\ahmin$,
but decreases somewhat as $\Phmin$ increases.  The terrace width, $R$, decreases as
$\Phmin$ increases, as it does with increases in $\ahmin$.  

\begin{figure}
\centering
\epsfig{file=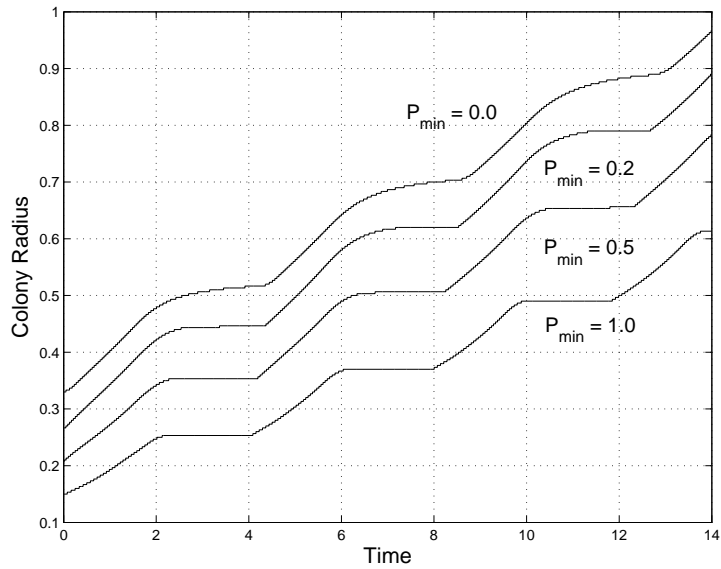,height=3in} 
\caption{Swarm-colony front dynamics as a function of $\Phmin$.  Shown
  are curves for the cases of $\Phmin = 0.0$, 
  $\Phmin = 0.2$, $\Phmin = 0.5$, and $\Phmin = 1.0$.
  The other parameters for these swarm colonies are $\ahmax = 2.67$, $V_c = 8$,
  $D_0 = 2\times 10^{-3}$, $\ahmin = 0.0$, and $\xi_0 = 0.5$.  The
  curves have been realigned in time for better
  comparison of individual ring formation.  This figure illustrates
  the loss of a distinct transition from swarming to consolidation
  phase as $\Phmin$ goes to zero.} 
\label{Pmin}
\end{figure}

The parameter $\Phmin$ is needed in the models to generate a distinct
consolidation phase.  The loss of a clear distinction between swarming
and consolidation as $\Phmin$ goes to zero is further
illustrated in Figure \ref{Pmin}. 

\begin{table}
\begin{center}
\begin{tabular}{|r||r|r|r|r|r|r|}
\hline
$\ahmax$ & $T$ & $S$ & $C$ & $S/C$ & $R$ & $R/S$ \\ \hline
2.00&3.8&1.5&2.3&0.65&0.090&0.0600\\ \hline
2.33&3.9&1.8&2.1&0.86&0.120&0.0667\\ \hline
{\bf 2.67}&4.1&2.2&1.9&1.16&0.153&0.0695\\ \hline
3.00&4.25&2.5&1.75&1.43&0.190&0.0760\\ \hline
3.50&4.5&3.2&1.3&2.46&0.253&0.0791\\ \hline
\end{tabular}
\caption{Ring formation as a function of $\ahmax$.}
\label{ahmax}
\end{center}
\end{table}

Table \ref{ahmax} exhibits ring formation as a function of $\ahmax$, the
dimensionless age at which a multinuclear swarmer cell breaks down
into mononuclear dividing cells.  A swarm colony with larger swarmer cells, i.e. larger $\ahmax$, can be expected to have
longer swarming phases, $S$, with wider terrace widths, $R$.  The
total cycle time, $T$, is also seen to increase as $\ahmax$ increases,
whereas the time spent in consolidation, $C$, decreases. 


\section{Conclusions}
\label{conclusions}

In this paper we have presented models and computational results to
understand the periodic front dynamics in {\em Proteus mirabilis}
swarm-colony development.  These models differ from previous work by
placing the most emphasis on known biological mechanisms, in
particular the local kinetics, or age-structure, and make fewer
assumptions about the diffusivity.  The numerical solutions to the
model show the periodicity seen in the biological system, as well as
the invariance of the cycle time to parameters sensitive to changes in
agar and glucose concentration, and retain a mechanism for
controlling the ratio of swarm to consolidation time within an
invariant total cycle time.  These results were obtained by robust and
accurate numerical methods designed specifically for age-structured
diffusion problems, with proven convergence properties.

Future work on further understanding {\em Proteus} swarm-colony
  development includes experimental studies on the forms of the 
  differentiation function, $\xi$.  Experiments to determine how differentiation from dividing cells to
  swarmer cells depends on local population density would
  be particularly valuable.

  Other areas of future research are the specifics of how the much
  larger and broader issue of biological motion relates to {\em Proteus} motion on an agar surface, and modeling the
  effects of a non-sharp age of septation on the regularity of   
  {\em Proteus} front dynamics.  Extending the models to
  explicit two-dimensional space to understand the interaction of
  multiple colonies is also an area of future research.

{\em Proteus mirabilis} should be seen as a model system with
relevance to other biological problems.  The further
mathematical understanding of age and space structured models -- and
the numerics to solve them -- constitutes an important foundational area of research.

\section*{Acknowledgements}
The author thanks James Shapiro at the University of Chicago for
  introducing him to {\em Proteus mirabilis}.  The author has had many
  helpful discussions with Todd F. Dupont, Tasso Kaper, Nancy 
  Kopell, Mitsugu Matsushita, Georgiy Medvedev, Thomas Nagylaki, and Oliver
  Rauprich.

\appendix

\section{Numerical Solutions of the Esipov-Shapiro Model}
\label{appendixSergei}
The Esipov-Shapiro model, as described by the relevant equations and
parameters in \cite{EnS}, has no spatial dynamics whatsoever.

We make several natural changes in the Esipov-Shapiro model to
generate something resembling {\em Proteus mirabilis} swarm-colony
development.  These changes, although not contained in their equations, are suggested in the text.

We set the units of the parameter $D_0$ to $\mbox{cm}^2$/hr rather than
$\mbox{cm}^2$/sec.  Use of the latter causes very rapid swarming that covers
the entire domain in a short time.  It is reasonable to believe
that this is a typo, as all other parameters in \cite{EnS} use hours as
the time unit.  The second change is the use of swarmer biomass rather
than numerical density as the relevant measure of total population.
Although the formal mathematical definition of $P_s$ (page 255 of
\cite{EnS}) is given by $P_s =
\int_{\theta_{\mbox{\tiny min}}}^{\theta_{\mbox{\tiny max}}} d\theta \,
\rho_s(\mathbf{r},\theta,t)$, the statement at the top of page 253,
\begin{quote}
Surface densities of swimmer and swarmer cell populations will be
indicated by the capital letters $P_c$ and $P_s$, respectively ($c$
for consolidation phase, $s$ for swarming phase).  While we do not
resolve swimmers in age, the surface density of swarmers is related to
a biomass weighted average over the ages, with $e^{\theta/\tau_d}$
being the contribution of age $\theta$.
\end{quote}
\noindent along with personal communication with Sergei Esipov (July,
1997), and the statement by the authors in \cite{rauprich} that their models in \cite{EnS} use an ``age-weighted swarmer cell
density'', 
indicates that a mass-weighted $P_s$ was used in their numerical
computations.  This is a necessary reinterpretation.  The use of a non-weighted $P_s$ precludes the upper
trigger of the memory field needed to instigate swarming,
$P_{s,\mbox{\small max}}$, from ever being reached, which in turn results in no
swarm-colony development whatsoever.

\begin{figure}[t]
\centering
\epsfig{file=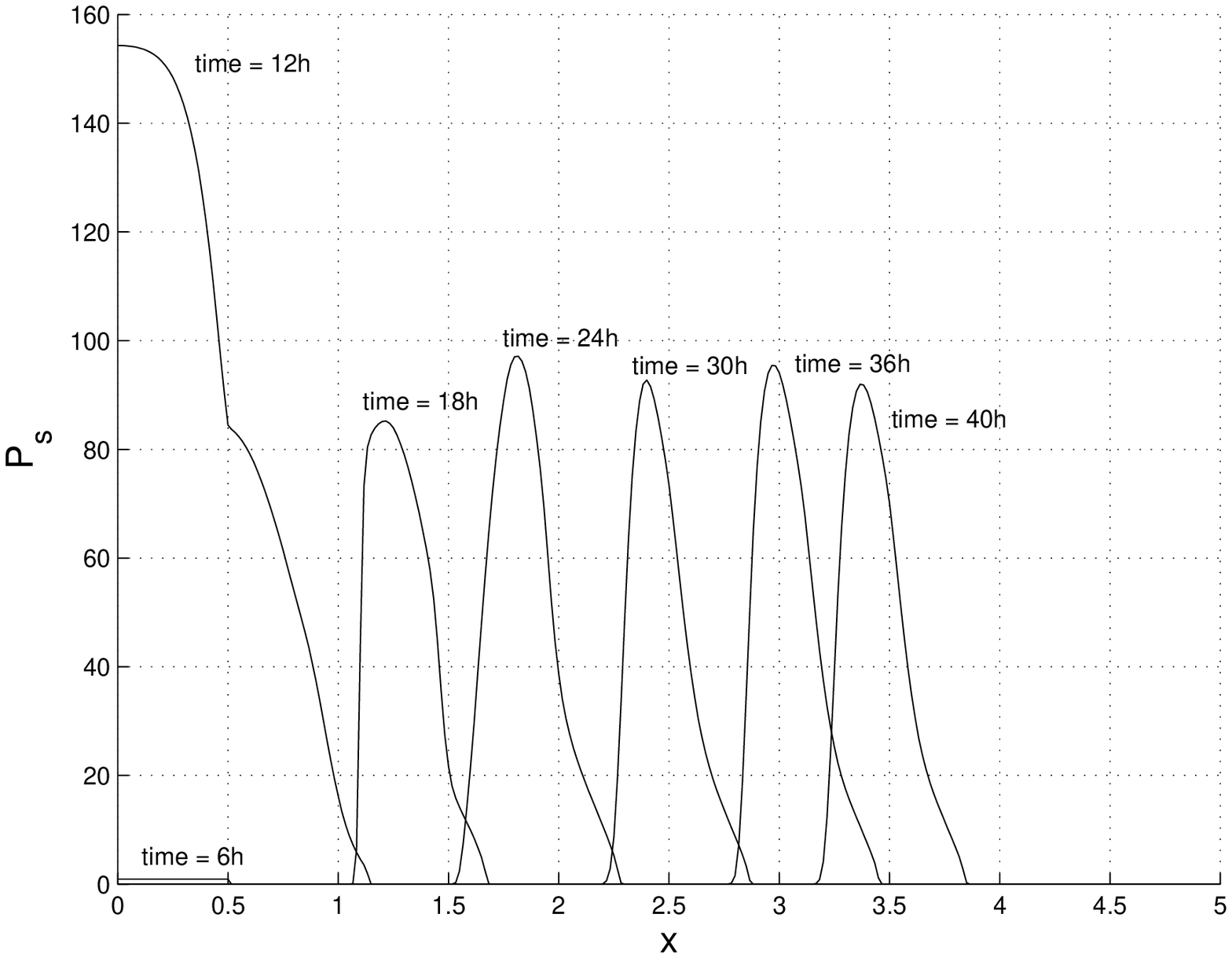,height=1.65in} \epsfig{file=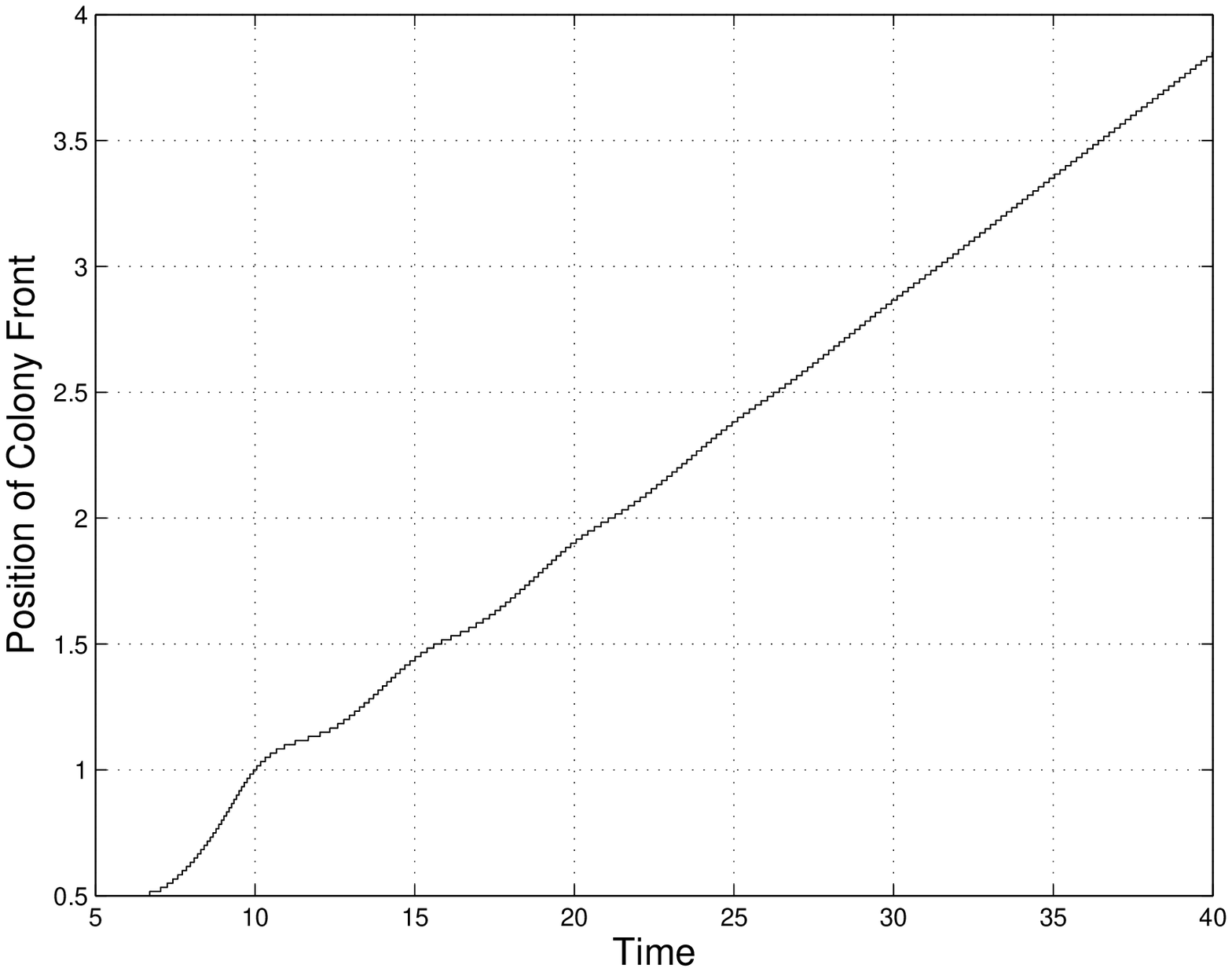,height=1.65in} 
\caption{Profiles of swarmer biomass for the Esipov-Shapiro model
  without $\Theta(\theta_{\mbox{\small min}})$ in the diffusivity. The model does not
  develop a consolidation phase after the first swarm phase, but
  rather the swarmers develop a self-sustaining soliton where swarmers
  of maximum age that dedifferentiate into dividing cells immediately
  differentiate into new swarmers.  The parameters
  used in this computation are taken from the caption of Figure 4 in
  \cite{EnS}, with the choice of $D_0$ = 0.045.}
\label{soliton}
\end{figure}

Another change in the Esipov-Shapiro model needed to get some sort of swarming and
consolidation behavior is the incorporation of what they refer to as
the ``drivers license'' age, $\theta_{\mbox{\small min}}$, explicitly
into the diffusivity.   As the equations
are written, $\theta_{\mbox{\small min}}$ enters the system implicitly
only through the
definition of $P_s$.  Thus, small swarmers move with larger swarmers in the ``rafts''; they just do
not contribute to a raft's motility.  This is a reasonable
assumption -- nothing prevents them from being caught up in the flow
-- and one made in the models presented in this 
paper. But because we use a lag phase in septation, the presence or
absence of ``underage'' swarmers in rafts does not qualitatively
affect our model. However, since it does not contain such a lag, the Esipov-Shapiro
model does not develop a consolidation phase after the first swarm
phase, but rather the swarmers develop a self-sustaining soliton
caused by 
swarmers of maximum age that dedifferentiate into swimmers (dividing
cells,) immediately differentiating into new swarmers.  Although this does
not reflect the observed behavior of {\em Proteus}, it is an
interesting mathematical object, which we display in Figure \ref{soliton}.

\begin{figure}[t]
\centering
\epsfig{file=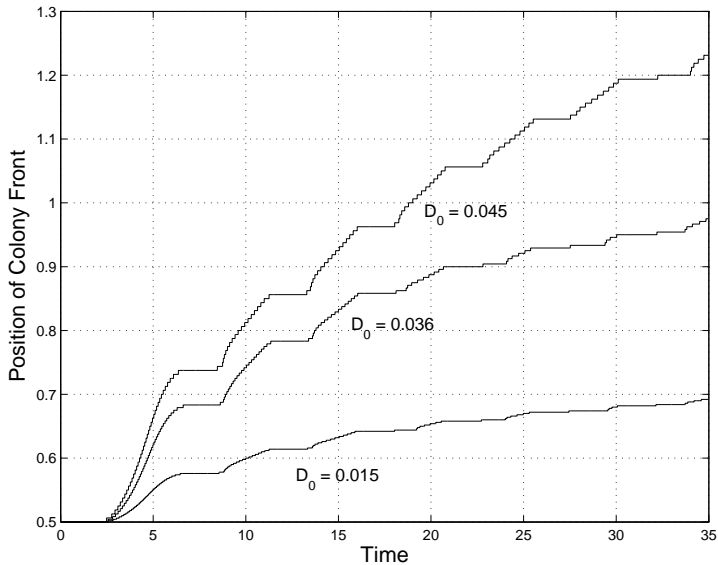,height=3in} 
\caption{Swarm-colony front dynamics of the
  Esipov-Shapiro model as a function of their $D_0$.  The parameters
  used in these computations are taken from the caption of Figure 4 in
  \cite{EnS}, including the choice of $D_0$ = (0.045, 0.036, 0.015).}
\label{damping}
\end{figure}

If we reinterpret the meaning of the ``drivers license'' age to be that
small swarmers do not move at all, we need to multiply the
diffusivity, $D$, as written in Equation (4a) on page 256 of
\cite{EnS}, by the Heaviside function $\Theta(\theta_{\mbox{\small min}})$.  Doing so results in a
system with a swarming/consolidation cycle with fixed time lengths,
but not the regularly spaced terraces seen in the experiments.  The terrace widths decrease in
length each cycle.  This is illustrated for several values of $D_0$ in
the Esipov-Shapiro model in Figure \ref{damping}.

The cause of the spatial regularity seen in the computations in
\cite{EnS} is quite possibly due to computational effects.  The
numerical integration in age and time in \cite{EnS} is not done along
characteristics, but rather consists of transforming age to a
stage-structured variable dependent upon discretization parameters,
adding 
regularity to a system that is not really there.  Another
possible source of error is that the space and time integration, as written in
Equations (6b)-(6d) on page 257 of \cite{EnS}, has the property that
the colony front can move at most one spatial interval per time step.
Thus, inaccurate numerics can change the nature of the bacterial
front. 

\section{Use of Different Forms of $\xi$}
\label{appendixXi}

We chose a smooth piecewise cubic function for the form of $\xi$
in Equation (\ref{diff}) for numerical reasons.  In this appendix, we
argue that it is the lag in $\xi$, not the shape of the curve, that is
relevant in modeling {\em Proteus} swarm-colony
development.  We present two sets of results from using two different piecewise constant
functions with the same area under the curve as the piecewise cubic.
The first, ``fat'' one,
\begin{equation}
 \xi_f(v) =  \left\{ \begin{array}{ll} \frac{1}{2}\xi_0, &
                      |v-v_c| \leq v_w, \\
                      0, & \mbox{otherwise,}
                      \end{array} \right.  \\ \label{fat}
\end{equation} 
has the same support, but shorter
maximum height, than the cubic.  The second, ``skinny'' one,
\begin{equation}
 \xi_s(v) =  \left\{ \begin{array}{ll} \xi_0, &
                      |v-v_c| \leq \frac{1}{2}v_w, \\
                      0, & \mbox{otherwise,}
                      \end{array} \right.  \\ \label{skinny}
\end{equation} 
has the same maximum height, but narrower support, than
the cubic.  Using Equation (\ref{fat}) instead of Equation
(\ref{diff}) gives Table \ref{VcFat} in place of the same data in Table
\ref{Vc}. Using Equation (\ref{skinny}) instead of Equation
(\ref{diff}) gives Table \ref{VcSkinny} in place of those data in Table
\ref{Vc}.  

The temporal and spatial regularity seen in {\em Proteus}
and the computations presented earlier with a piecewise cubic $\xi$ are
also present in these computations.  These results indicate that changing the form of $\xi$ does not alter
the qualitative behavior of the system and has only a minor
quantitative effect on the solutions.  Nonetheless, the shape of
$\xi$, and not only its mass and support, is an interesting topic for experiments.

\begin{table}
\begin{center}
\begin{tabular}{|r||r|r|r|r|r|r|}
\hline
$V_c$ & $T$ & $S$ & $C$ & $R$ \\ \hline
2   &3.3&2.1&1.2&0.097 \\ \hline
4&3.7&2.1&1.6&0.120 \\ \hline
{\bf 8}&4.1&2.1&2.0&0.131 \\ \hline
16&4.7&2.1&2.6&0.134 \\ \hline
\end{tabular}
\caption{Ring formation as a function of $V_c$ using Equation (\ref{fat}).}
\label{VcFat}
\end{center}
\end{table}

\begin{table}
\begin{center}
\begin{tabular}{|r||r|r|r|r|r|r|}
\hline
$V_c$ & $T$ & $S$ & $C$ & $R$ \\ \hline
2   &3.2&2.2&1.0&0.117  \\ \hline
4&3.5&2.3&1.2&0.147 \\ \hline
{\bf 8}&4.0&2.3&1.7&0.164 \\ \hline
16&4.5&2.3&2.2&0.176 \\ \hline
\end{tabular}
\caption{Ring formation as a function of $V_c$ using Equation (\ref{skinny}).}
\label{VcSkinny}
\end{center}
\end{table}

\bibliographystyle{siam}
\bibliography{age,na,math,bio,forest,spatialecology}

\end{document}

%% file: trcover.tex
%
%
%
%

\voffset -.25in

\thispagestyle{empty}
\fontsize{12}{12}
\setcounter{page}{0}

\begin{center}
  \centerline{
  \psfig{figure=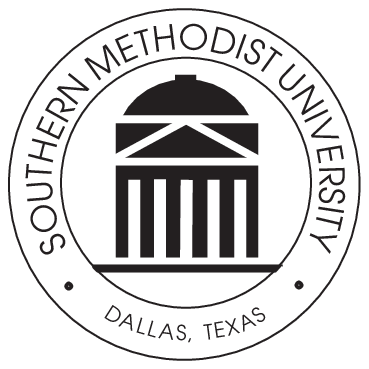,height=1.5in,width=1.5in}
  }

  \vspace{1.25in}

  \begin{minipage}[t]{4in}
    \begin{center}
  
      {\bf \trtitle

      \vspace{24pt}

      \trauths

      \vspace{12pt}

      SMU Math Report \trnum 
      }
    \end{center}
  \end{minipage}

  \vspace{3in}

  {\Large\bf D}{\large\bf EPARTMENT OF} 
  {\Large\bf M}{\large\bf ATHEMATICS} 

  \vspace{3pt}

  {\Large\bf S}{\large\bf OUTHERN} 
  {\Large\bf M}{\large\bf ETHODIST} 
  {\Large\bf U}{\large\bf NIVERSITY} 
\end{center}

\voffset 0in

\normalsize
\newpage

%% file: techRep.bbl
\begin{thebibliography}{10}

\bibitem{aronson85}
{\sc D.~G. Aronson}, {\em The role of diffusion in mathematical population
  biology: {Skellam} revisited}, in Mathematics in Biology and Medicine,
  V.~Capasso, E.~Grosso, and S.~L. Paveri-Fontana, eds., vol.~57 of Lecture
  Notes in Biomathematics, Springer-Verlag, Berlin, 1985, pp.~2--6.

\bibitem{age-pwconst-paper}
{\sc B.~P. Ayati}, {\em A variable time step method for an age-dependent
  population model with nonlinear diffusion}, SIAM J. Numer. Anal., 37 (2000),
  pp.~1571--1589.

\bibitem{age-general-paper}
{\sc B.~P. Ayati and T.~F. Dupont}, {\em Galerkin methods in age and space for
  a population model with nonlinear diffusion}, SIAM J. Numer. Anal., 40
  (2002), pp.~1064--1076.

\bibitem{step-doubling-paper}
\leavevmode\vrule height 2pt depth -1.6pt width 23pt, {\em Convergence of a
  step-doubling \uppercase{G}alerkin method for parabolic problems}, Math.
  Comp., posted electronically September 10, 2004, to appear in print. (2004).

\bibitem{cancerSurvey}
{\sc N.~Bellomo and L.~Preziosi}, {\em Modelling and mathematical problems
  related to tumor evolution and its interaction with the immune system},
  Mathematical and Computer Modelling, 32 (2000), pp.~413--452.

\bibitem{BertuzziGandolfi}
{\sc A.~Bertuzzi and A.~Gandolfi}, {\em Cell kinetics in a tumor cord}, J.
  Theo. Biol., 204 (2000), pp.~587--599.

\bibitem{BolkerNPacalaNClaudia}
{\sc B.~M. Bolker, S.~W. Pacala, and C.~Neuhauser}, {\em Spatial dynamics in
  model plant communities: What do we really know?}, American Naturalist, 162
  (2003), pp.~135--148.

\bibitem{BnI}
{\sc S.~Busenberg and M.~Iannelli}, {\em A class of nonlinear diffusion
  problems in age-dependent population dynamics}, Nonlin. Anal. Th. Meth.
  Appl., 7 (1983), pp.~501--529.

\bibitem{chiu}
{\sc C.~Chiu}, {\em A numerical method for nonlinear age dependent population
  models}, Diff. Int. Eqns., 3 (1990), pp.~767--782.

\bibitem{deRoos89}
{\sc A.~M. {de Roos}}, {\em Numerical methods for structured population models:
  The escalator boxcar train}, Num. Meth. Part. Diff. Eqns., 4 (1989),
  pp.~173--195.

\bibitem{deRoos97}
\leavevmode\vrule height 2pt depth -1.6pt width 23pt, {\em A gentle
  introduction to physiologically structured population models}, in
  Structured-population Models in Marine, Terrestrial, and Freshwater Systems,
  S.~Tuljapurkar and H.~Caswell, eds., vol.~18 of Population and Community
  Biology Series, Chapman \& Hall, New York, 1997, ch.~5, pp.~119--204.

\bibitem{deRoos91}
{\sc A.~M. {de Roos} and J.~A.~J. Metz}, {\em Towards a numerical analysis of
  the escalator boxcar train}, in Differential Equations with Applications in
  Biology, Physics and Engineering, J.~A. Goldstein, F.~Kappel, and
  W.~Schappacher, eds., vol.~133 of Lecture Notes in Pure and Applied
  Mathematics, Marcel Dekker, New York, 1991, pp.~91--113.

\bibitem{diblasio}
{\sc G.~{di Blasio}}, {\em Non-linear age-dependent population diffusion}, J.
  Math. Bio., 8 (1979), pp.~265--284.

\bibitem{DnL}
{\sc G.~{di Blasio} and L.~Lamberti}, {\em An initial-boundary problem for
  age-dependent population diffusion}, SIAM J. Appl. Math., 35 (1978),
  pp.~593--615.

\bibitem{DysonWebb}
{\sc J.~Dyson, R.~Villella-Bressan, and G.~F. Webb}, {\em Asynchronous
  exponential growth in an age structured population of proliferating and
  quiescent cells}, Math. Biosci., 177-178 (2002), pp.~73--83.

\bibitem{EnS}
{\sc S.~E. Esipov and J.~A. Shapiro}, {\em Kinetic model of {\em
  \uppercase{p}roteus mirabilis\/} swarm colony development}, J. Math. Biol.,
  36 (1998), pp.~249--268.

\bibitem{FnL-M}
{\sc G.~Fairweather and J.~C. L{\'o}pez-Marcos}, {\em A box method for a
  nonlinear equation of population dynamics}, IMA J. Numer. Anal., 11 (1991),
  pp.~525--538.

\bibitem{gear}
{\sc C.~W. Gear}, {\em Numerical Initial Value Problems in Ordinary
  Differential Equations}, Prentice--Hall, New Jersey, 1971.

\bibitem{gurtin}
{\sc M.~E. Gurtin}, {\em A system of equations for age-dependent population
  diffusion}, J. Theor. Biol., 40 (1973), pp.~389--392.

\bibitem{GnM1}
{\sc M.~E. Gurtin and R.~C. MacCamy}, {\em Non-linear age-dependent population
  dynamics}, Arch. Rat. Mech. Anal., 54 (1974), pp.~281--300.

\bibitem{GnM2}
\leavevmode\vrule height 2pt depth -1.6pt width 23pt, {\em Diffusion models for
  age-structured populations}, Math. Biosci., 54 (1981), pp.~49--59.

\bibitem{MainiDicty}
{\sc T.~H\"{o}fer, J.~A. Sherratt, and P.~K. Maini}, {\em Cellular pattern
  formation during {\em \uppercase{d}ictostelium} aggregation}, Physica D, 85
  (1995), pp.~425--444.

\bibitem{huang}
{\sc C.~Huang}, {\em An age-dependent population model with nonlinear diffusion
  in $\mathbb{R}^n$}, Quart. Appl. Math., 52 (1994), pp.~377--398.

\bibitem{IonidesStochastic}
{\sc E.~L. Ionides, K.~S. Fang, R.~R. Isseroff, and G.~F. Oster}, {\em
  Stochastic models for cell motion and taxis}, J. Math. Biol., 48 (2004),
  pp.~23--37.

\bibitem{kim}
{\sc M.-Y. Kim}, {\em Galerkin methods for a model of population dynamics with
  nonlinear diffusion}, Num. Meth. Part. Diff. Eqns., 12 (1996), pp.~59--73.

\bibitem{KnP}
{\sc M.-Y. Kim and E.-J. Park}, {\em Mixed approximation of a population
  diffusion equation}, Computers Math. Applic., 30 (1995), pp.~23--33.

\bibitem{Kohyama92FE}
{\sc T.~Kohyama}, {\em Size-structured multi-species model of rain forest
  trees}, Functional Ecology, 6 (1992), pp.~206--212.

\bibitem{Kohyama93}
\leavevmode\vrule height 2pt depth -1.6pt width 23pt, {\em Size-structured tree
  populations in gap-dynamic forest -- the forest architecture hypothesis for
  the stable coexistence of species}, J. Ecology, 81 (1993), pp.~131--143.

\bibitem{KooiNKooijman}
{\sc B.~W. Kooi and S.~A. L.~M. Kooijman}, {\em Discrete event versus
  continuous approach to reproduction in structured population dynamics}, Theo.
  Popul. Biol., 56 (1999), pp.~91--105.

\bibitem{Kalachev}
{\sc M.~A. Kraemer, L.~V. Kalachev, and D.~W. Coble}, {\em A class of models
  describing age structure dynamics in a natural forest}, Natural Resource
  Modeling, 15 (2002), pp.~149--200.

\bibitem{KnL}
{\sc K.~Kubo and M.~Langlais}, {\em Periodic solutions for a population
  dynamics problem with age-dependence and spatial structure}, J. Math. Biol.,
  29 (1991), pp.~363--378.

\bibitem{KnC}
{\sc Y.~Kwon and C.-K. Cho}, {\em Second-order accurate difference methods for
  a one-sex model of population dynamics}, SIAM J. Numer. Anal., 30 (1993),
  pp.~1385--1399.

\bibitem{langlais1}
{\sc M.~Langlais}, {\em A nonlinear problem in age-dependent population
  diffusion}, SIAM J. Math. Anal., 16 (1985), pp.~510--529.

\bibitem{langlais2}
\leavevmode\vrule height 2pt depth -1.6pt width 23pt, {\em Large time behavior
  in a nonlinear age-dependent population dynamics problem with spatial
  diffusion}, J. Math. Biol., 26 (1988), pp.~319--346.

\bibitem{LnT}
{\sc L.~Lopez and D.~Trigiante}, {\em A finite difference scheme for a stiff
  problem arising in the numerical solution of a population dynamic model with
  spatial diffusion}, Nonlin. Anal. Th. Meth. Appl., 9 (1985), pp.~1--12.

\bibitem{L-M}
{\sc J.~C. L{\'o}pez-Marcos}, {\em An upwind scheme for a nonlinear hyperbolic
  integro-differential equation with integral boundary condition}, Computers
  Math. Applic., 22 (1991), pp.~15--28.

\bibitem{maccamy}
{\sc R.~C. MacCamy}, {\em A population model with nonlinear diffusion}, J.
  Diff. Eqns., 39 (1981), pp.~52--72.

\bibitem{marcati}
{\sc P.~Marcati}, {\em Asymptotic behavior in age-dependent population dynamics
  with hereditary renewal law}, SIAM J. Math. Anal., 12 (1981), pp.~904--916.

\bibitem{MnKnK}
{\sc G.~S. Medvedev, T.~J. Kaper, and N.~Kopell}, {\em A reaction-diffusion
  equation with periodic front dynamics}, SIAM J. Appl. Math., 60 (2000),
  pp.~1601--1638.

\bibitem{Murray3-2}
{\sc J.~D. Murray}, {\em Mathematical Biology II: Spatial Models and Biomedical
  Applications}, vol.~18 of Interdisciplinary Applied Mathematics,
  Springer--Verlag, Berlin, third~ed., 2003.

\bibitem{nagy89}
{\sc T.~Nagylaki}, {\em The diffusion model for migration and selection}, in
  Some Mathematical Questions in Biology: Models in Population Biology,
  A.~Hastings, ed., vol.~20 of Lectures on Mathematics in the Life Sciences,
  American Mathematical Society, Providence, R.I., 1989, pp.~55--75.

\bibitem{PascualNLevin}
{\sc M.~Pascual and S.~A. Levin}, {\em Spatial scaling in a benthic population
  model with density-dependent disturbance}, Theor. Popul. Biol., 56 (1999),
  pp.~106--122.

\bibitem{cancerModelling}
{\sc L.~Preziosi}, ed., {\em Cancer Modelling and Simulation}, Chapman \&
  Hall/CRC, 2003.

\bibitem{rauprich}
{\sc O.~Rauprich, M.~Matsushita, K.~Weijer, F.~Siegert, S.~E. Esipov, and J.~A.
  Shapiro}, {\em Periodic phenomena in {\em \uppercase{p}roteus mirabilis\/}
  swarm colony development}, J. Bacteriol., 178 (1996), pp.~6525--6538.

\bibitem{roten}
{\sc M.~Rotenberg}, {\em Theory of population transport}, J. Theor. Biol., 37
  (1972), pp.~291--305.

\bibitem{skellam}
{\sc J.~G. Skellam}, {\em Random dispersal in theoretical populations},
  Biometrika, 38 (1951), pp.~196--218.

\bibitem{sulsky}
{\sc D.~Sulsky}, {\em Numerical solution of structured population models,
  \uppercase{I}. age structure}, J. Math. Biol., 31 (1993), pp.~817--839.

\bibitem{Webb}
{\sc G.~F. Webb}, {\em Theory of Nonlinear Age--dependent Population Dynamics},
  vol.~89 of Pure and Applied Mathematics, Marcel Dekker, New York, 1985.

\bibitem{WnSproteus}
{\sc F.~D. Williams and R.~H. Schwarzhoff}, {\em Nature of the swarming
  phenomenon in {\em \uppercase{p}roteus}}, Ann. Rev. Microbiol., 32 (1978),
  pp.~101--122.

\end{thebibliography}
